\documentclass{ws-procs975x65}
\usepackage{amssymb}
\usepackage{graphicx}

\begin{document}

\title{Central compact objects in magnetic lethargy.}
\author{Vigan\`o, Daniele$^1$, Pons, Jose A.$^1$, Perna, Rosalba$^2$}
\address{$^1$Departament de F\'{\i}sica Aplicada, Universitat d'Alacant\\ Ap. Correus 99, 03080 San Vicent del Raspeig, Spain\\
$^2$ Department of Astrophysical and Planetary Sciences and JILA, University of Colorado\\ 440 UCB Boulder, 80309, USA\\
E-mail: daniele.vigano@ua.es}

\def \aj {Astronomical Journal}
\def \ATel {Astron.\ Tel.}
\def \apj {ApJ}
\def \apjl {ApJL}
\def \apjs {ApJS}
\def \aap {A\&A}
\def \aapr {A\&ARv.}
\def \aara {A\&A Rev.}
\def \apss {AP\&SS}
\def \araa {Annu. Rev. Astro. Astrophys.}
\def \cjaa {Chin. J. Astron. Astrophys.} 
\def \mnras {MNRAS}
\def \nat {Nat} 
\def \physrep {Phys. Rep.}
\def \pasj {PASJ}
\def \prd {Phys. Rev. D}
\def \rmxaa {RMAA}
\def \solphys {Solar Physics}
\def \curlB {\vec{\nabla}\times (e^\nu \vec{B})}
\newcommand{\dive} {\vec{\nabla}\cdot}
\newcommand{\rot} {\vec{\nabla}\times}

\begin{abstract}
Central Compact Objects are peculiar young neutron stars, with very low external magnetic fields combined with high fluxes in the
X-ray band and surface temperature anisotropies. However, in their crust the magnetic field can be strong, result of its burial during a short post-supernova hypercritical accretion episode. The implications of this latter scenario for the temperature anisotropy, 
pulsed fraction and luminosity are discussed.
\end{abstract}
%

\bodymatter\bigskip

Among the heterogenous family of isolated neutron stars\cite{kaspi10} (NSs), there are $\sim$ 10 so-called ``Central Compact Objects'' (CCOs)\cite{gotthelf13}, because of their location close to the center of detected supernova remnants (SNRs). Their distinguishing properties are the lack of radio, optical and $\gamma$ emission, a thermal spectrum in the X-ray band, low inferred magnetic field (MF), and young age. The latter can be estimated by kinematic studies of associated SNRs. Timing properties are known only for the three NSs
in the SNRs kes79, Puppis A and G296.5+10.0. In all three cases, unusually low values of the spin period derivative are measured, $\dot{P}\sim 10^{-17}-10^{-18} s\,s^{-1}$, implying that the observed periods, $0.1-0.4$ s, are very close to the birth values, and that the magnetic torque is very low. The inferred dipolar fields for the three cases are in the range $B\sim 10^{10}-10^{11}$ G, low enough for these objects to deserve the label as ``anti-magnetars``. This picture of apparently magnetically dead neutron stars is also consistent with the absence of radio-activity, although selection effects due to beaming cannot be excluded, given the low statistics.

However, in several sources, X-ray data suggest a different scenario inside these sources. In particular, the two hot-spots of the CCO in Puppis A\cite{deluca12}, and the very high pulsed fraction of the CCO in Kes79 (evidence for a large surface temperature anisotropy) are inconsistent within the picture of a low MF\cite{shabaltas12}, as meridional temperature gradients caused by anisotropic thermal conduction are only evident for $B\gtrsim 10^{13}$ G. 
From a theoretical point of view, detailed simulations of MHD equilibrium configurations\cite{ciolfi09}, show that a non-negligible fraction of magnetic energy can be initially stored in the interior, in the form of toroidal and multipolar components. In addition, the state-of-the-art of the magneto-thermal evolution models\cite{vigano12a}, accounting for Ohmic and Hall terms in the induction equation, show that during the first $10^3-10^5$ years, Hall dynamics leads to the formation of multipolar and toroidal components inside the crust, not 
necessarily reflected by the external, dipolar field.

Hence, the ``hidden magnetic field scenario'' \cite{young95,muslimov95,geppert99} as a viable model for CCOs has become more 
popular \cite{ho11}, challenging and complementing the ``anti-magnetar'' scenario. Here we briefly 
discuss the main results from the first 2D simulations of submergence and reemergence of the field, considering consistently the thermal and magnetic evolution. We refer to our recent paper\cite{vigano12b} for more technical details.

During a time interval of few months after the supernova explosion, the newly born NS accretes material from the reversed shock, at a super-Eddington rate\cite{colgate71,chevalier89}. This hypercritical accretion could bury the MF into the NS crust, at a depth strongly dependent on the amount of matter and on geometry of accretion, resulting in an external MF (responsible for the spindown of the star) much lower than the internal ``hidden" MF.  We found that nearly isotropic accretion of $M\gtrsim  10^{-4} M_\odot$ is able to reduce by orders of magnitude the initial external field, by the formation of screening currents in the crust. The MF inside the star, however, can locally reach very large values. When accretion stops, the screening currents in the outer crust are dissipated and the magnetic field slowly reemerges. 
When the reemergence has completed, the external field reaches a value slightly smaller than the original one, due to the dissipation. The timescale of the reemergence strongly depends on the amount of accreted material\cite{geppert99,ho11,vigano12b}, ranging from thousands to millions of years.

\begin{figure}[t]
 \centering
\includegraphics[height=.4\textwidth]{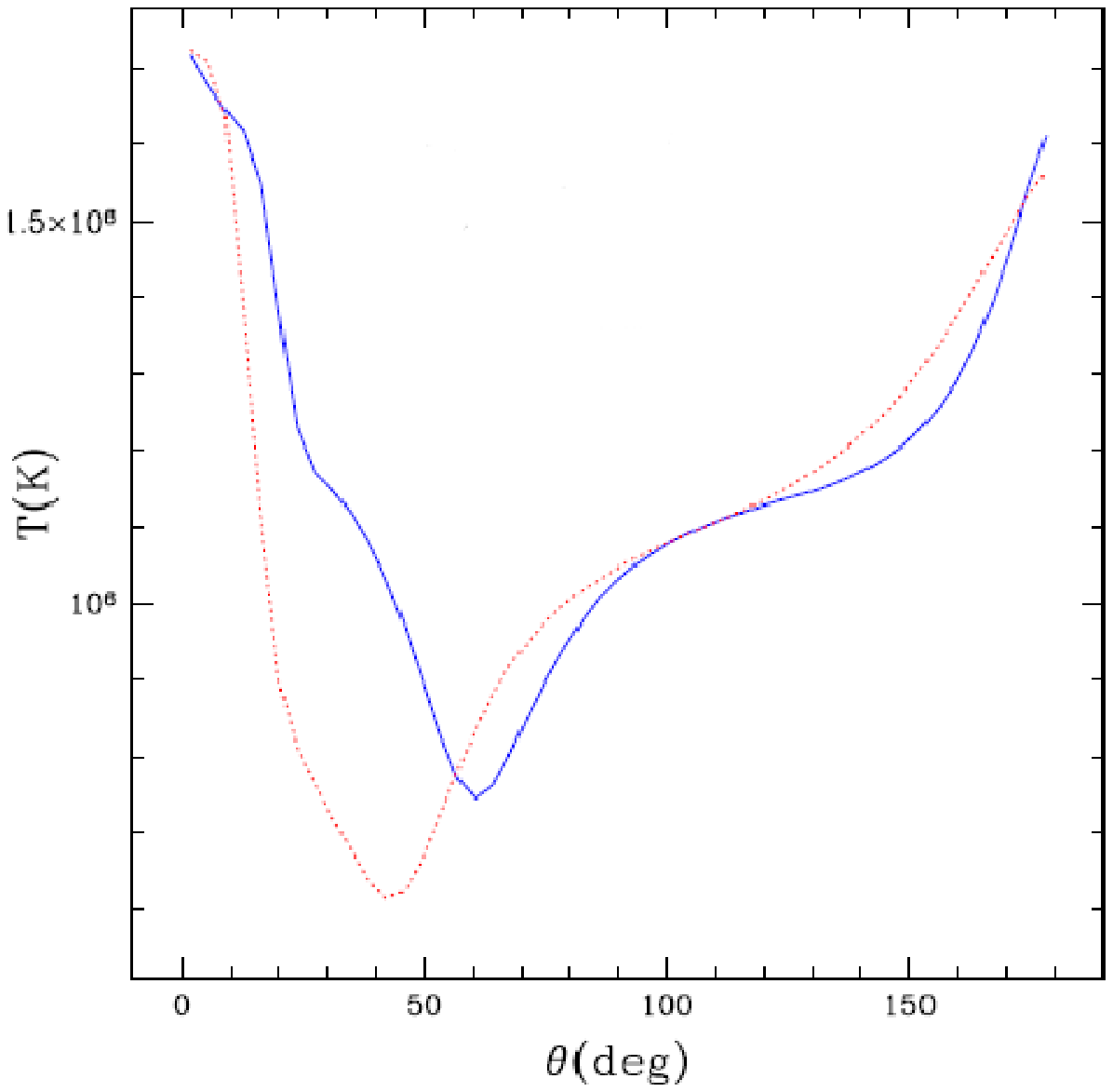}
\includegraphics[height=.4\textwidth,width=.5\textwidth]{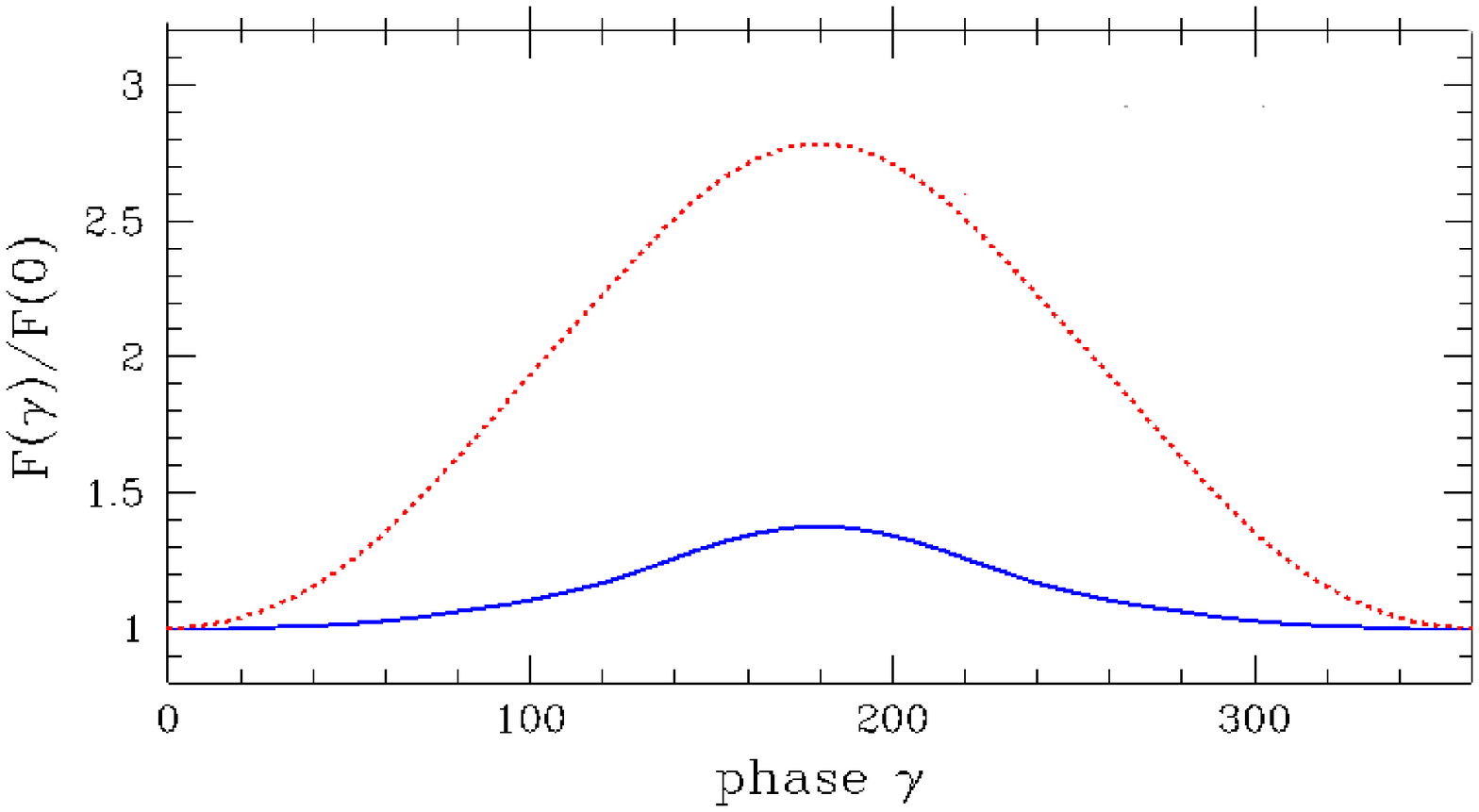}
\caption{Surface temperature and relative light curve for an initial model with $10^{13}$ G dipolar poloidal field, $10^{15}$ G toroidal dipolar field, during two different times of the reemergence stage.} 
 \label{fig:tcco}
\end{figure}

While the observable effects on timing properties have been deeply discussed\cite{ho11,vigano12b}, little has been done
to explain the surface temperature distribution and luminosity.
In Fig.~\ref{fig:tcco} we present the surface temperature profile and the resulting light curve, during the reemergence phase, in the range 0.5-2 keV calculated with the general relativistic effects\cite{perna12}. The initial model is quite extreme, consisting in a $10^{13}$ G dipolar poloidal field, with a $10^{15}$ G toroidal dipolar field, and a post-SN accretion of $10^{-3} M_\odot$. During the reemergence phase, the toroidal field acts as an insulator and drifts towards one pole due to the Hall term in the induction equation. This results in a large anisotropic temperature
distribution consisting in a large cold region and a small hot spot in the north pole. Although the problem is highly degenerated, this is an example that can quantitatively explain the exceptional pulsed fraction observed for kes79 (see dotted red line on the right panel), confirming the proposed existence of a strong 
interior toroidal field\cite{shabaltas12}. Models with more standard values of initial magnetic field, or at different stages of reemergence
(like blue solid line) produce little anisotropy. The diversity of CCOs reflects in this sense the different geometries of the
hidden MF at a given age.

Moreover, CCOs are on average more luminous than young RPPs\cite{vigano13}, and this can be due to two effects. First, if the internal magnetic field is strong enough, the energy provided by the dissipation of screening currents can increase the luminosity. We note that this strongly depends on the model, in particular on the depth of the screening currents. Only in the case in which currents are submerged to shallow depths, the luminosity can be increased by a factor of a few. Second, due to the initial accretion, the envelope of these NSs is likely to consist of light elements, which provide at early times larger luminosities than Iron envelopes\cite{vigano13}. 

Last, we point out to a natural link between RPPs and CCOs. When the reemergence stage is over, the external field recovers a standard value, and most likely the associated SNR has faded away: a CCO can turn into a RPP. If the accreted mass is not very high, the field is submerged to shallow depths, and the reemergence stage lasts $10^3$-$10^4$ years, during which the braking index will be slightly less than 3\cite{pons12}. This is in agreement with the reported values measured for young pulsars ($<10^4$ yrs) \cite{hobbs10,espinoza11}.

\section*{Acknowledgments}
This work was partly supported by the Spanish grant AYA 2010-21097-C03-02 and CompStar, a Research Networking Program of the ESF. DV is supported by a fellowship from the \textit{Prometeo} program for research groups of excellence of the Generalitat Valenciana (Prometeo/2009/103). DV thanks J.~Halpern for helpful discussions.

\label{lastpage}
\bibliographystyle{ws-procs975x65}
\bibliography{cco}

\end{document}